\pdfoutput=1

\documentclass[12pt,a4paper]{article}

\usepackage{ifthen} 
\newboolean{pdflatex}
\setboolean{pdflatex}{true} 

\newboolean{articletitles}
\setboolean{articletitles}{true} 

\newboolean{uprightparticles}
\setboolean{uprightparticles}{false} 

\usepackage{amssymb}
\usepackage{amsfonts}
\usepackage{upgreek} 
\usepackage{subfig}
\usepackage{pdflscape}
\newsubfloat{figure}
\newsubfloat{table}

\usepackage[abs]{overpic}
\usepackage{verbatim}
\usepackage{nicefrac}

\usepackage{hyperref}

\usepackage{microtype}
\usepackage{lineno}  
\usepackage{xspace} 

\usepackage{graphicx}  
\usepackage{color}
\usepackage{colortbl}

\usepackage{amsmath} 
\usepackage{amssymb}
\usepackage{amsfonts}
\usepackage{upgreek} 

\newcommand*\patchAmsMathEnvironmentForLineno[1]{%
\expandafter\let\csname old#1\expandafter\endcsname\csname #1\endcsname
\expandafter\let\csname oldend#1\expandafter\endcsname\csname
end#1\endcsname
 \renewenvironment{#1}%
   {\linenomath\csname old#1\endcsname}%
   {\csname oldend#1\endcsname\endlinenomath}%
}
\newcommand*\patchBothAmsMathEnvironmentsForLineno[1]{%
  \patchAmsMathEnvironmentForLineno{#1}%
  \patchAmsMathEnvironmentForLineno{#1*}%
}
\AtBeginDocument{%
\patchBothAmsMathEnvironmentsForLineno{equation}%
\patchBothAmsMathEnvironmentsForLineno{align}%
\patchBothAmsMathEnvironmentsForLineno{flalign}%
\patchBothAmsMathEnvironmentsForLineno{alignat}%
\patchBothAmsMathEnvironmentsForLineno{gather}%
\patchBothAmsMathEnvironmentsForLineno{multline}%
}

\usepackage{hyperref}    
\usepackage[all]{hypcap} 

\def\lhcb {\mbox{LHCb}\xspace}
\def\ux85 {\mbox{UX85}\xspace}

\def\babar  {\mbox{BaBar}\xspace}
\def\belle  {\mbox{Belle}\xspace}

\ifthenelse{\boolean{uprightparticles}}%
{

 \def\PDelta      {\ensuremath{\Delta}\xspace}                 
 \def\PXi      {\ensuremath{\Xi}\xspace}                 
 \def\PLambda      {\ensuremath{\Lambda}\xspace}                 
 \def\PSigma      {\ensuremath{\Sigma}\xspace}                 
 \def\POmega      {\ensuremath{\Omega}\xspace}                 
 \def\PUpsilon      {\ensuremath{\Upsilon}\xspace}                 
 
 \def\PB      {\ensuremath{\mathrm{B}}\xspace}                 
                  
 \def\PD      {\ensuremath{\mathrm{D}}\xspace}

 \def\PK      {\ensuremath{\mathrm{K}}\xspace}

 \def\Pb      {\ensuremath{\mathrm{b}}\xspace}                 
 \def\Pc      {\ensuremath{\mathrm{c}}\xspace}

 \def\Pi      {\ensuremath{\mathrm{i}}\xspace}

}
{

 \mathchardef\PDelta="7101
 \mathchardef\PXi="7104
 \mathchardef\PLambda="7103
 \mathchardef\PSigma="7106
 \mathchardef\POmega="710A
 \mathchardef\PUpsilon="7107
                  
 \def\PB      {\ensuremath{B}\xspace}                 
                  
 \def\PD      {\ensuremath{D}\xspace}

 \def\PK      {\ensuremath{K}\xspace}

 \def\Pb      {\ensuremath{b}\xspace}                 
 \def\Pc      {\ensuremath{c}\xspace}

 \def\Pi      {\ensuremath{i}\xspace}

}

\def\cquark    {\ensuremath{\Pc}\xspace}

\def\bquark    {\ensuremath{\Pb}\xspace}

\def\kaon  {\ensuremath{\PK}\xspace}
  \def\Kbar  {\kern 0.2em\overline{\kern -0.2em \PK}{}\xspace}

\def\Kz    {\ensuremath{\kaon^0}\xspace}
\def\Kzb   {\ensuremath{\Kbar^0}\xspace}
\def\KzKzb {\ensuremath{\Kz \kern -0.16em \Kzb}\xspace}
\def\Kp    {\ensuremath{\kaon^+}\xspace}
\def\Km    {\ensuremath{\kaon^-}\xspace}

\def\KpKm  {\ensuremath{\Kp \kern -0.16em \Km}\xspace}
 
\def\KL    {\ensuremath{\kaon^0_{\rm\scriptscriptstyle L}}\xspace}

  \def\Dbar    {\kern 0.2em\overline{\kern -0.2em \PD}{}\xspace}
\def\D       {\ensuremath{\PD}\xspace}

\def\Dz      {\ensuremath{\D^0}\xspace}
\def\Dzb     {\ensuremath{\Dbar^0}\xspace}
\def\DzDzb   {\ensuremath{\Dz {\kern -0.16em \Dzb}}\xspace}
\def\Dp      {\ensuremath{\D^+}\xspace}
\def\Dm      {\ensuremath{\D^-}\xspace}

\def\DpDm    {\ensuremath{\Dp {\kern -0.16em \Dm}}\xspace}

  \def\Bbar    {\kern 0.18em\overline{\kern -0.18em \PB}{}\xspace}

  \def\Y#1S{\ensuremath{\PUpsilon{(#1S)}}\xspace}

\def\Lbar {\ensuremath{\kern 0.1em\overline{\kern -0.1em\PLambda}}\xspace}

\def\BF         {{\ensuremath{\cal B}\xspace}}

\def\ra                 {\ensuremath{\rightarrow}\xspace}
\def\to                 {\ensuremath{\rightarrow}\xspace}

\def\AT#1     {\ensuremath{A_{\mathrm{T}}^{#1}}\xspace}           

\def\C#1      {\ensuremath{\mathcal{C}_{#1}}\xspace}                       
\def\Cp#1     {\ensuremath{\mathcal{C}_{#1}^{'}}\xspace}                    
\def\Ceff#1   {\ensuremath{\mathcal{C}_{#1}^{\mathrm{(eff)}}}\xspace}        
\def\Cpeff#1  {\ensuremath{\mathcal{C}_{#1}^{'\mathrm{(eff)}}}\xspace}       
\def\Ope#1    {\ensuremath{\mathcal{O}_{#1}}\xspace}                       
\def\Opep#1   {\ensuremath{\mathcal{O}_{#1}^{'}}\xspace}                    

\newcommand{\tev}{\ensuremath{\mathrm{\,Te\kern -0.1em V}}\xspace}
\newcommand{\gev}{\ensuremath{\mathrm{\,Ge\kern -0.1em V}}\xspace}
\newcommand{\mev}{\ensuremath{\mathrm{\,Me\kern -0.1em V}}\xspace}
\newcommand{\kev}{\ensuremath{\mathrm{\,ke\kern -0.1em V}}\xspace}
\newcommand{\ev}{\ensuremath{\mathrm{\,e\kern -0.1em V}}\xspace}
\newcommand{\gevc}{\ensuremath{{\mathrm{\,Ge\kern -0.1em V\!/}c}}\xspace}
\newcommand{\mevc}{\ensuremath{{\mathrm{\,Me\kern -0.1em V\!/}c}}\xspace}
\newcommand{\gevcc}{\ensuremath{{\mathrm{\,Ge\kern -0.1em V\!/}c^2}}\xspace}
\newcommand{\gevgevcccc}{\ensuremath{{\mathrm{\,Ge\kern -0.1em V^2\!/}c^4}}\xspace}
\newcommand{\mevcc}{\ensuremath{{\mathrm{\,Me\kern -0.1em V\!/}c^2}}\xspace}

\def\mum  {\ensuremath{\,\upmu\rm m}\xspace}

\def\invfb   {\ensuremath{\mbox{\,fb}^{-1}}\xspace}

\def\gsim{{~\raise.15em\hbox{$>$}\kern-.85em
          \lower.35em\hbox{$\sim$}~}\xspace}
\def\lsim{{~\raise.15em\hbox{$<$}\kern-.85em
          \lower.35em\hbox{$\sim$}~}\xspace}

\def\ptot       {\mbox{$p$}\xspace}
\def\pt         {\mbox{$p_{\rm T}$}\xspace}

\def\evtgen     {\mbox{\textsc{EvtGen}}\xspace}
\def\pythia     {\mbox{\textsc{Pythia}}\xspace}

\def\geant      {\mbox{\textsc{Geant4}}\xspace}

\def\photos     {\mbox{\textsc{Photos}}\xspace}

\def\tell1  {TELL1\xspace}
\def\ukl1   {UKL1\xspace}

\newcommand{\CLs}{\ensuremath{\textrm{CL}_{\textrm{s}}}\xspace}

\newcommand{\BuJpsimmK}{\ensuremath{B^-\to J/\psi(\mu^+\mu^-) K^-}\xspace}

\newcommand{\Jpsimumu}{\ensuremath{J/\psi\to \mu^+\mu^-}\xspace}

\newcommand{\tmmm}{\ensuremath{\tau^-\to \mu^-\mu^+\mu^-}\xspace}

\newcommand{\DsPhiPi}{\ensuremath{D_s^-\to \phi \left(\mu^+\mu^-\right) \pi^-}\xspace}
\newcommand{\DsPhiKKPi}{\ensuremath{D_s^-\to \phi \left(K^+K^-\right) \pi^-}\xspace}
\newcommand{\DsTauNu}{\ensuremath{D_s^-\to \tau^- \bar{\nu}_{\tau}}\xspace}
 
\newcommand{\Phimm}{\ensuremath{\phi\to \mu^+\mu^-}\xspace} 
\newcommand{\PhiKK}{\ensuremath{\phi\to K^+K^-}\xspace} 
\newcommand{\DsEtaMuNu}{\ensuremath{\D_s^- \to \eta \left(\mu^+\mu^-\gamma\right) \mu^- {\bar{\nu}_\mu}\xspace}}

\newcommand{\BRof}[1]{\ensuremath{{\cal B}\left(#1\right)}\xspace}
\newcommand{\gl}{\ensuremath{{\rm \mathcal{M}_{3body}}}\xspace}
\newcommand{\pid}{\ensuremath{{\rm \mathcal{M}_{PID}}}\xspace}

\usepackage{cite} 
\usepackage{mciteplus}
\usepackage{booktabs}

\begin{document}

\renewcommand{\thefootnote}{\fnsymbol{footnote}}
\setcounter{footnote}{1}


\begin{titlepage}
\pagenumbering{roman}

\vspace*{-1.5cm}
\centerline{\large EUROPEAN ORGANIZATION FOR NUCLEAR RESEARCH (CERN)}
\vspace*{1.5cm}
\hspace*{-0.5cm}
\begin{tabular*}{\linewidth}{lc@{\extracolsep{\fill}}r}
\ifthenelse{\boolean{pdflatex}}
{\vspace*{-2.7cm}\mbox{\!\!\!\includegraphics[width=.14\textwidth]{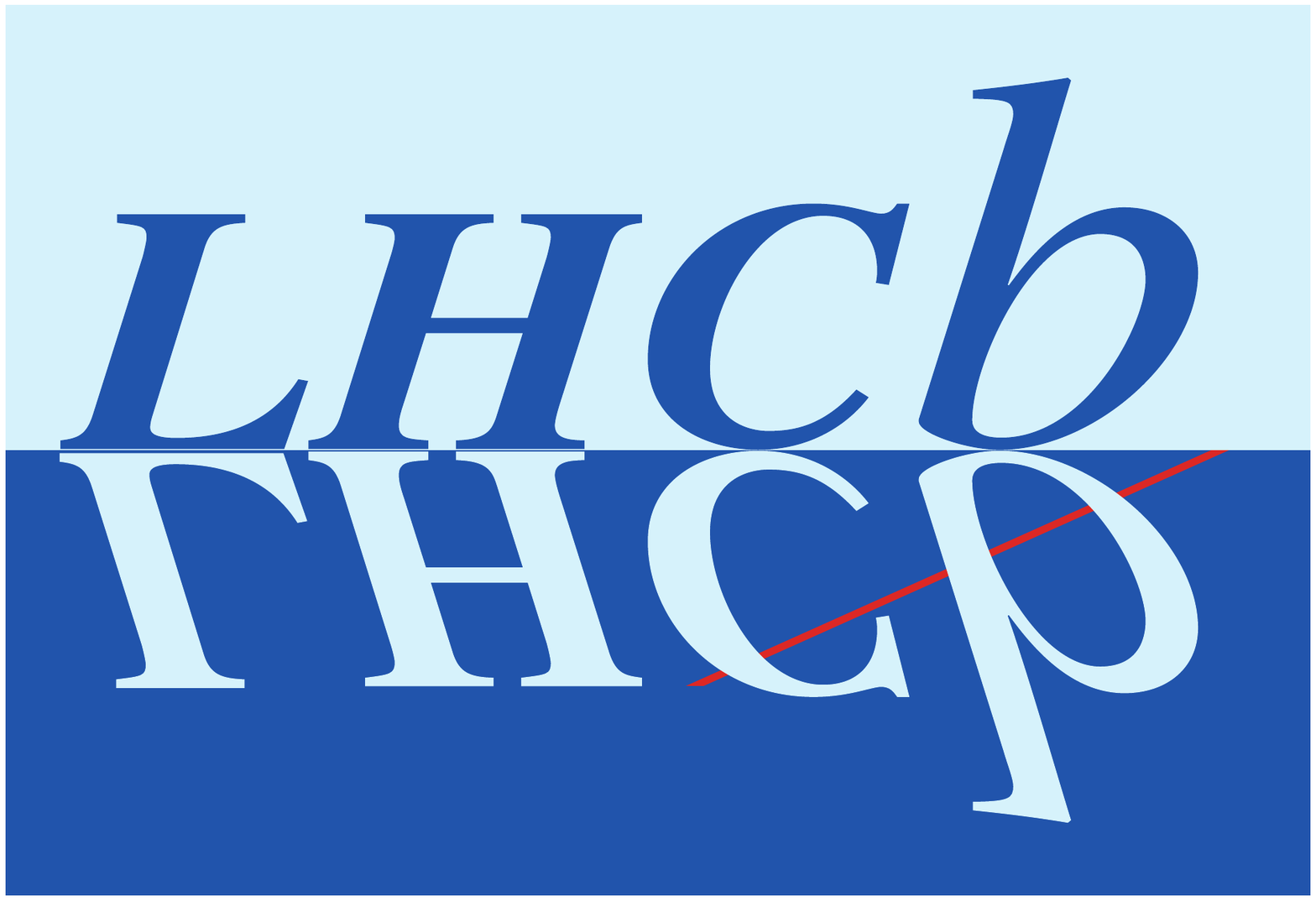}} & &}%
{\vspace*{-1.2cm}\mbox{\!\!\!\includegraphics[width=.12\textwidth]{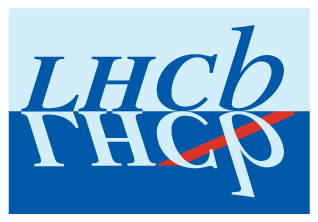}} & &}%
\\
 & & CERN-PH-EP-2014-236 \\  
 & & LHCb-PAPER-2014-052 \\  
 & & February 18, 2015 \\ 
 & & \\
\end{tabular*}

\vspace*{4.0cm}

{\bf\boldmath\huge
\begin{center}
  Search for the lepton flavour violating decay \tmmm
\end{center}
}

\vspace*{2.0cm}

\begin{center}
The LHCb collaboration\footnote{Authors are listed at the end of this paper.}
\end{center}

\vspace{\fill}

\begin{abstract}
  \noindent
A search for the lepton flavour violating decay \tmmm is performed with the LHCb experiment.
The data sample corresponds to an integrated luminosity of $1.0$\invfb of proton-proton 
collisions at a centre-of-mass energy of 7\tev and $2.0$\invfb at 8\tev. 
No evidence is found for a signal, and a limit 
is set at $90\%$ confidence level on the branching fraction, $\BF(\tmmm) < 4.6 \times 10^{-8}$.
\end{abstract}

\vspace*{1.0cm}

\begin{center}
  Published as JHEP 02 (2015) 121
\end{center}

\vspace{\fill}

{\footnotesize 
\centerline{\copyright~CERN on behalf of the \lhcb collaboration, licence \href{http://creativecommons.org/licenses/by/4.0/}{CC-BY-4.0}.}}
\vspace*{2mm}

\end{titlepage}


\newpage
\setcounter{page}{2}
\mbox{~}
\newpage


\cleardoublepage

\renewcommand{\thefootnote}{\arabic{footnote}}
\setcounter{footnote}{0}



\pagestyle{plain} 
\setcounter{page}{1}
\pagenumbering{arabic}


%

\section{Introduction}
\label{sec:Introduction}

Lepton flavour violating processes are allowed within the context of the Standard Model (SM) with 
massive neutrinos, but their branching 
fractions are of order $10^{-40}$~\cite{Raidal:2008jk, Ilakovac:2012sh} or smaller, 
and are beyond the reach of any currently conceivable experiment.
Observation of charged lepton flavour violation (LFV) would therefore be an unambiguous signature 
of physics beyond the Standard Model (BSM), but no such process 
has been observed to date~\cite{PDG2014}. 

A number of BSM scenarios predict LFV at branching fractions 
approaching current experimental sensitivities~\cite{lfvreview}, with LFV in $\tau^-$ 
decays
often enhanced with respect to $\mu^-$ decays due to the large difference in mass between the two leptons 
(the inclusion of charge-conjugate processes is implied throughout). 
If charged LFV were to be discovered, measurements of the branching fractions for a number of channels 
would be required to determine the nature of the BSM physics. 
In the absence of such a discovery, improving the experimental constraints on the branching fractions for LFV decays would help to constrain 
the parameter spaces of BSM models.

This paper reports on an updated search for the LFV decay
\tmmm with the \lhcb experiment~\cite{Alves:2008zz} at the CERN LHC. 
The previous \lhcb analysis of this channel produced the first result on a search for LFV $\tau^-$ decays at a 
hadron collider~\cite{paper1fb}.  
Using $1.0$\invfb of proton-proton collision data collected at a centre-of-mass energy of 7\tev,
a limit was set on the branching fraction, $\BF\left(\tmmm\right) < 8.0 \times 10^{-8}$ at 90\% confidence level (CL).
The current best experimental upper limits are 
$\BF\left(\tmmm\right)<2.1 \times 10^{-8}$ at 90\% CL from \belle~\cite{Hayasaka:2010np}
and $\BF\left(\tmmm\right)<3.3 \times 10^{-8}$ at 90\% CL from \babar~\cite{Lees:2010ez}.
In the analysis presented here, an additional \lhcb data set, corresponding to $2.0$\invfb of integrated luminosity collected at 
8\tev, is added to the previous data set, and a number of new analysis techniques are introduced. 

The search for LFV in $\tau^-$ decays at \lhcb takes advantage of the large inclusive $\tau^-$ production cross-section 
at the LHC, where
$\tau^-$ leptons are produced almost entirely from the decays of $b$ and $c$ hadrons. 
Using the $b\overline{b}$ and $c\overline{c}$ cross-sections measured by \lhcb~\cite{sigmabbLHCb,sigmaccLHCb} and the inclusive 
$b\rightarrow\tau$ and $c\rightarrow\tau$ branching fractions~\cite{PDG2014}, the inclusive $\tau^-$ 
cross-section is estimated to be $85\,\upmu$b at 7\tev.

Selection criteria are implemented for the signal mode, \tmmm, and for the
calibration and normalisation channel, which is $D_s^-\to\phi\pi^{-}$ with $\phi\to\mu^{+}\mu^{-}$, 
referred to in the following as \DsPhiPi.  
To avoid potential bias, $\mu^-\mu^+\mu^-$ candidates with mass
within $\pm 30\mevcc$~(approximately three times the expected mass resolution) of the known $\tau^-$ mass 
are initially excluded from the analysis.
Discrimination between a potential signal and the background 
is performed using a three-dimensional binned distribution in two multivariate classifiers and
the mass of the $\tau^-$ candidate. One classifier is based on the three-body decay topology 
and the other on muon identification. 

\section{Detector and triggers}
\label{sec:Detector}

The \lhcb detector~\cite{Alves:2008zz} is a single-arm forward
spectrometer covering the \mbox{pseudorapidity} range $2<\eta <5$,
designed for the study of particles containing \bquark or \cquark
quarks. The detector includes a high-precision tracking system
consisting of a silicon-strip vertex detector surrounding the $pp$
interaction region, a large-area silicon-strip detector located
upstream of a dipole magnet with a bending power of about
$4{\rm\,Tm}$, and three stations of silicon-strip detectors and straw
drift tubes placed downstream of the magnet.
The tracking system provides a measurement of momentum, \ptot,  with
a relative uncertainty that varies from 0.4\% at low momentum to 0.6\% at 100\gevc.
The minimum distance of a track to a primary vertex, the impact parameter (IP), is measured with a resolution of $\left(15+29/\pt\right)\mum$,
where \pt is the component of \ptot transverse to the beam, in \gevc.
Different types of charged hadrons are distinguished using information
from two ring-imaging Cherenkov detectors (RICH)~\cite{LHCb-DP-2012-003}. Photon, electron and
hadron candidates are identified by a calorimeter system consisting of
scintillating-pad and preshower detectors, an electromagnetic
calorimeter and a hadronic calorimeter. Muons are identified by a
system composed of alternating layers of iron and multiwire
proportional chambers~\cite{LHCb-DP-2012-002}. 

The trigger~\cite{trigger2011} consists of a
hardware stage, based on information from the calorimeter and muon
systems, followed by a software stage, which applies a full event
reconstruction.
Candidate events are first required to pass the hardware trigger, 
which selects muons with a transverse momentum $\pt>1.48\gevc$ in the 7\tev data 
or $\pt>1.76\gevc$ in the 8\tev data. 
In the software trigger, at least 
one of the final-state particles is required to have both 
$\pt>0.8\gevc$ and IP $>100\mum$ with respect to all 
of the primary $pp$ interaction vertices~(PVs) in the event. 
Finally, the tracks of two or more of the final-state 
particles are required to form a vertex that is significantly 
displaced from the PVs.

\section{Monte Carlo simulation}
\label{sec:Simulation}

In the simulation, $pp$ collisions are generated using
\pythia~\cite{Sjostrand:2006za,*Sjostrand:2007gs}
with a specific \lhcb configuration~\cite{LHCb-PROC-2010-056}. 
Decays of hadronic particles are described by \evtgen~\cite{Lange:2001uf}, 
in which final-state radiation is generated using \photos~\cite{Golonka:2005pn}. 
For the \tmmm signal channel, the final-state particles
are distributed according to three-body phase-space.
The interaction of the generated particles with the detector and its
response are implemented using the \geant
toolkit~\cite{Allison:2006ve, *Agostinelli:2002hh} as described in
Ref.~\cite{LHCb-PROC-2011-006}.

As the $\tau^-$ leptons produced in the LHCb acceptance originate almost exclusively from heavy quark decays, 
they can be classified in one of five categories according to the parent particle. 
The parent particle can be the following: a $b$ hadron; a $D_s^-$ or $D^-$ meson that is produced directly in 
a proton-proton collision or via the decay of an excited charm meson; 
or a $D_s^-$ or $D^-$ meson resulting from the decay of a $b$ hadron. 
Events from each category are generated separately and are combined in accordance with the measured 
cross-sections and branching fractions. 
Variations of the cross-sections and branching fractions within their uncertainties are considered 
as sources of systematic uncertainty.

\section{Event selection}
\label{sec:selection}

Candidate \tmmm decays are selected by requiring
three tracks that combine to give a mass close to that of 
the $\tau^-$ lepton, and that form 
a vertex that is displaced from the PV.
The tracks are required to be well-reconstructed muon candidates
with $\pt > 300$\,\mevc that have a significant separation
from the PV.
There must be a good fit to the three-track vertex,
and the decay time of
the candidate forming the vertex has to satisfy $ct > 100\mum$.
As the $\tau^-$ leptons are produced predominantly in the decays of charm mesons,
where the $Q$-values are relatively small (and so the charm meson and the $\tau^-$ are almost collinear in the laboratory frame), 
a requirement on the pointing angle, $\theta$, between 
the momentum vector of the three-track system and
the vector joining the primary and secondary vertices is used to remove poorly reconstructed candidates ($\cos\theta > 0.99$).  Contamination from pairs of tracks originating from the same
particle is reduced by removing same-sign muon pairs with mass lower than 250\mevcc.

The decay \DsEtaMuNu\, is a source of irreducible background near the signal region, 
and therefore candidates with a $\mu^+\mu^-$ invariant mass below $450$\,\mevcc are removed.
Signal candidates containing muons that result from the decay of the $\phi(1020)$ meson are removed by 
excluding $\mu^+\mu^-$ masses within $\pm 20\mevcc$ of the known $\phi(1020)$ meson mass. 

The signal region is defined by a $\pm 20\mevcc$ window~(approximately two times the expected mass resolution)
around the known $\tau^-$ mass. Candidates with $\mu^-\mu^+\mu^-$ 
invariant mass between 1600 and 1950\mevcc are kept to allow evaluation of the background 
contributions in the signal region. In the following, the wide mass windows on 
either side of the signal region are referred to as the data sidebands.
The signal region for the normalisation channel, \DsPhiPi,  which has a similar topology to that of the \tmmm decay,
is defined by a $\pm 20\mevcc$ window 
around the $D_s^-$ mass, 
with the $\mu^+\mu^-$ mass required to be within $\pm20\mevcc$ of the $\phi(1020)$ meson mass. 
Where appropriate, the rest of the selection criteria are identical to those for the signal channel, with one 
of the muon candidates replaced by a pion candidate. 

\section{Signal and background discrimination}
\label{sec:likelihoods}

Three classifiers are used to discriminate between signal and background: 
an invariant mass classifier that uses the 
reconstructed mass of the $\tau^-$ candidate;
a geometric classifier, \gl; and a particle identification classifier, \pid.

\begin{figure}[t]
\begin{minipage}[b]{0.5\linewidth}
        \centering
        \begin{overpic}[width=0.95\textwidth]{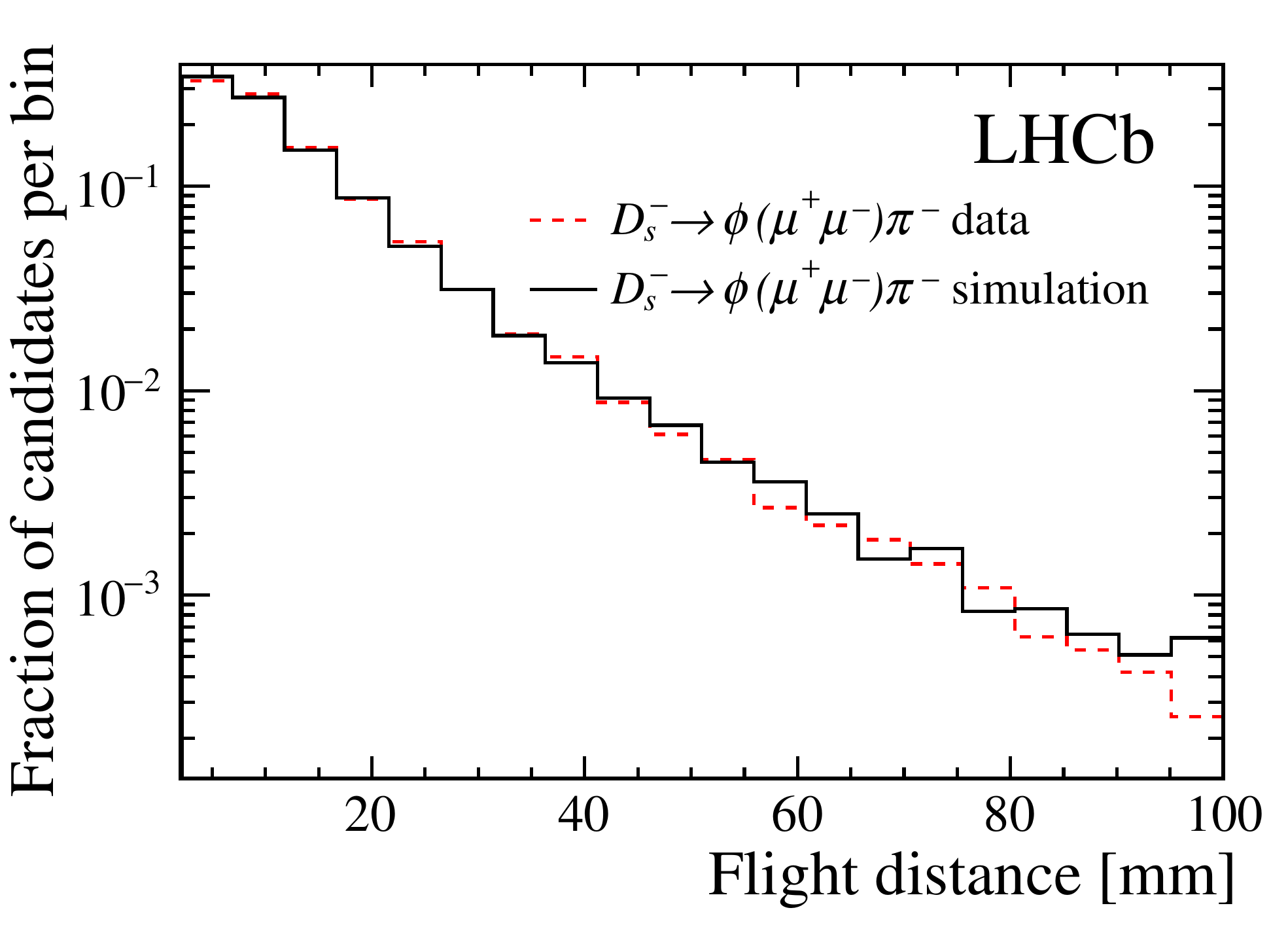}
        \put (40,100) {\small{(a)}}
        \end{overpic}
\end{minipage}
\begin{minipage}[b]{0.5\linewidth}
        \centering
        \begin{overpic}[width=0.95\textwidth]{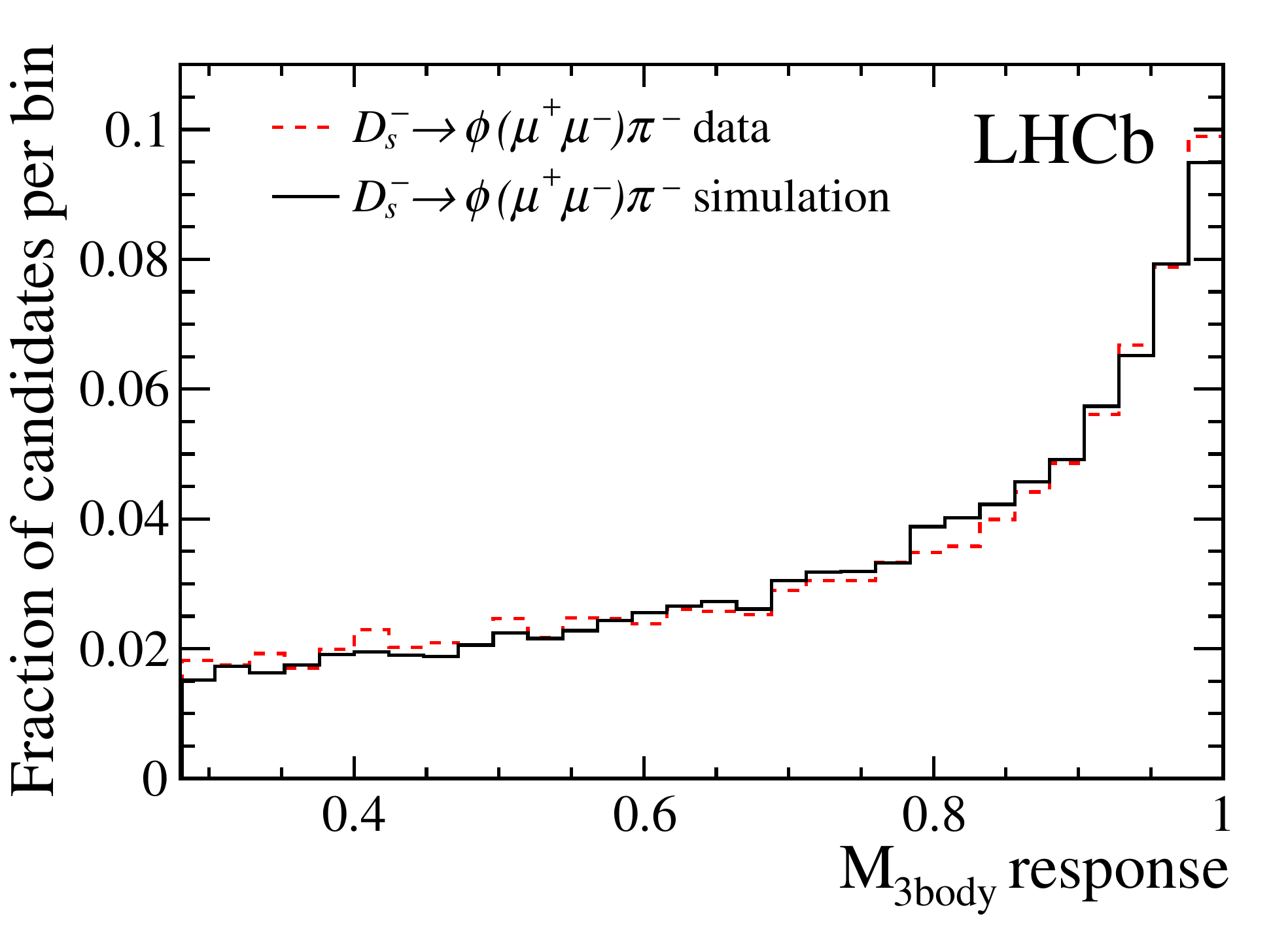}
        \put (40,100) {\small{(b)}}
        \end{overpic}
\end{minipage}

\caption{\small {Distribution of (a) $D_s^-$ flight distance and (b) \gl response for \DsPhiPi candidates at 8\tev. The dashed (red) lines indicate the data and the solid (black) lines indicate the simulation.
The data is background-subtracted using the sPlot technique~\cite{sweight}.
}}
\label{fig:dataMC}
\end{figure}

The multivariate classifier \gl is based on the geometry and kinematic properties 
of the final-state tracks and the reconstructed $\tau^-$ candidate. It aims to reject backgrounds 
from combinations of tracks that do not share a common vertex and those from multi-body decays 
with more than three final-state particles.
The variables used in the classifier include the vertex fit quality, the displacement of the vertex from the 
PV, the pointing angle $\theta$, and the IP and fit $\chi^2$ of the tracks. An ensemble-selected (blended)~\cite{blend}, 
custom boosted decision tree (BDT) classifier is used~\cite{MatrixNet, Breiman}, as described in the following. 
In the blending method the input variables are combined~\cite{tmva} into one BDT, 
two Fisher discriminants~\cite{fisher}, four neural networks~\cite{neuralNet}, one function-discriminant 
analysis~\cite{FDA} and one linear discriminant~\cite{LD}. 
Each classifier is trained using simulated signal and background samples, 
where the composition of the background is a mixture of 
$b\bar{b}\rightarrow\mu\mu X$ and $c\bar{c}\rightarrow\mu\mu X$
processes according to their relative abundances as measured in data. 
As each category of simulated signal events has different kinematic properties, a 
separate set of classifiers is trained for each. 
One third of the available signal sample is used 
at this stage, along with one half of the background sample.  
The classifier responses, along with the original input variables, are then used as input to the
custom BDT classifier, which is trained on the remaining half of the background sample 
and a third of the signal sample, with the five categories combined, to give the final classifier response.
The responses of the classifier on the training and the test samples are found to be in good agreement, 
suggesting no overtraining of the classifier is present.  
As the responses of the individual classifiers are not fully correlated, 
blending the output of the classifiers improves the sensitivity of the analysis in our data sample by 6\% with respect to that achievable 
by using the best single classifier. 
The \gl classifier response is calibrated using the
\DsPhiPi control channel to correct for differences in response between data and simulation.  
Figure~\ref{fig:dataMC} shows good agreement between \DsPhiPi data and simulation for one of the input variables to \gl 
and for the classifier response. 
A systematic uncertainty of 2\% is assigned to account for any remaining differences.
The classifier response is found to be uncorrelated with mass for both the 
signal sample and the data sidebands.

The multivariate classifier \pid uses information from the RICH detectors, 
the calorimeters and the muon detectors to obtain 
the likelihood that each of the three final-state particles is compatible with the muon hypothesis. 
The value of the \pid response is taken as the smallest likelihood of the three muon candidates. 
The \pid classifier uses a neural network that is trained on simulated events to discriminate muons from 
other charged particles. 
The \pid classifier response is calibrated using
muons from \Jpsimumu decays in data. 
 
\begin{figure}[t]
\begin{minipage}[b]{0.5\linewidth}
        \centering
        \begin{overpic}[width=0.95\textwidth]{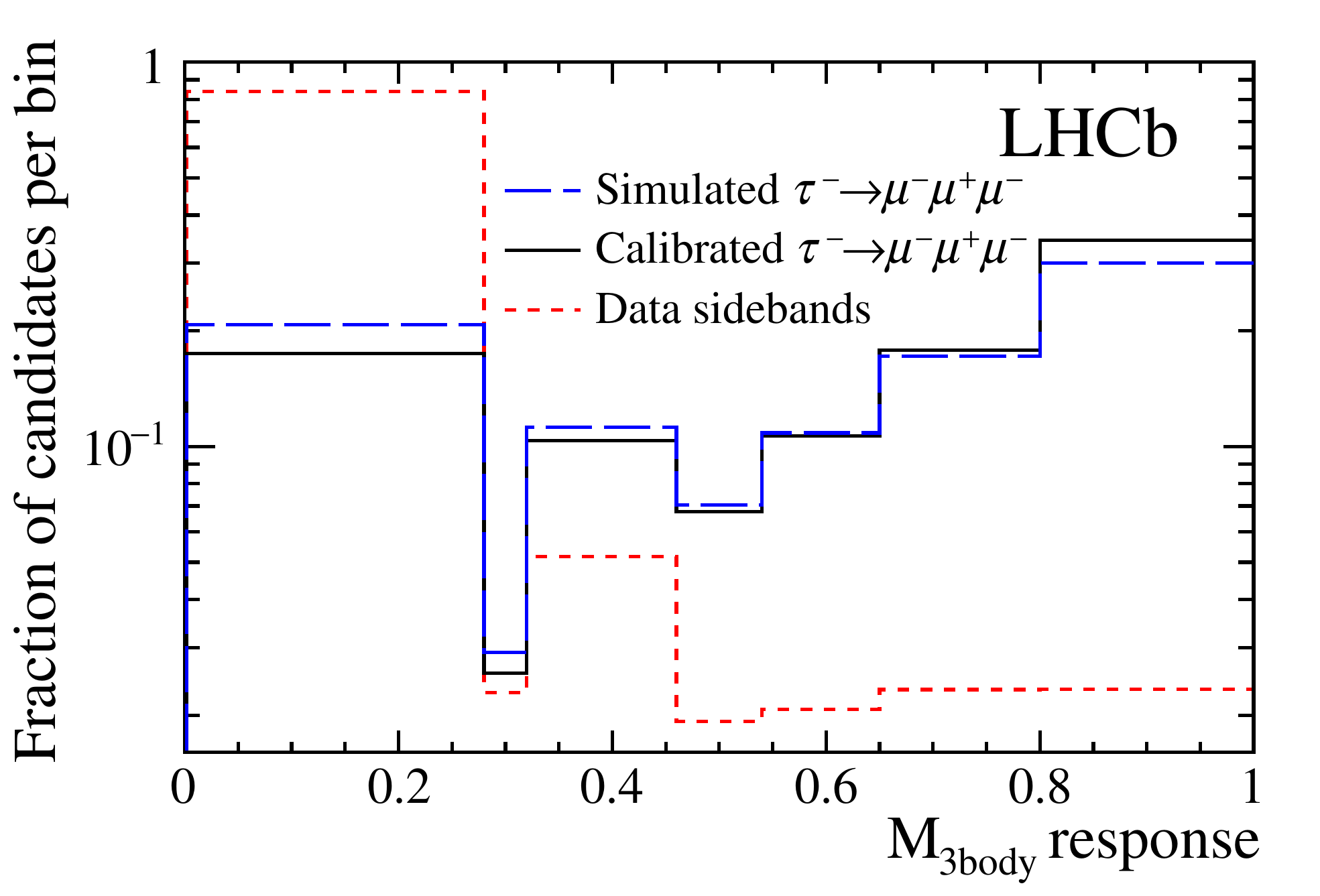}
        \put (40,115) {\small{(a)}}
        \end{overpic}
\end{minipage}
\begin{minipage}[b]{0.5\linewidth}
        \centering
        \begin{overpic}[width=0.95\textwidth]{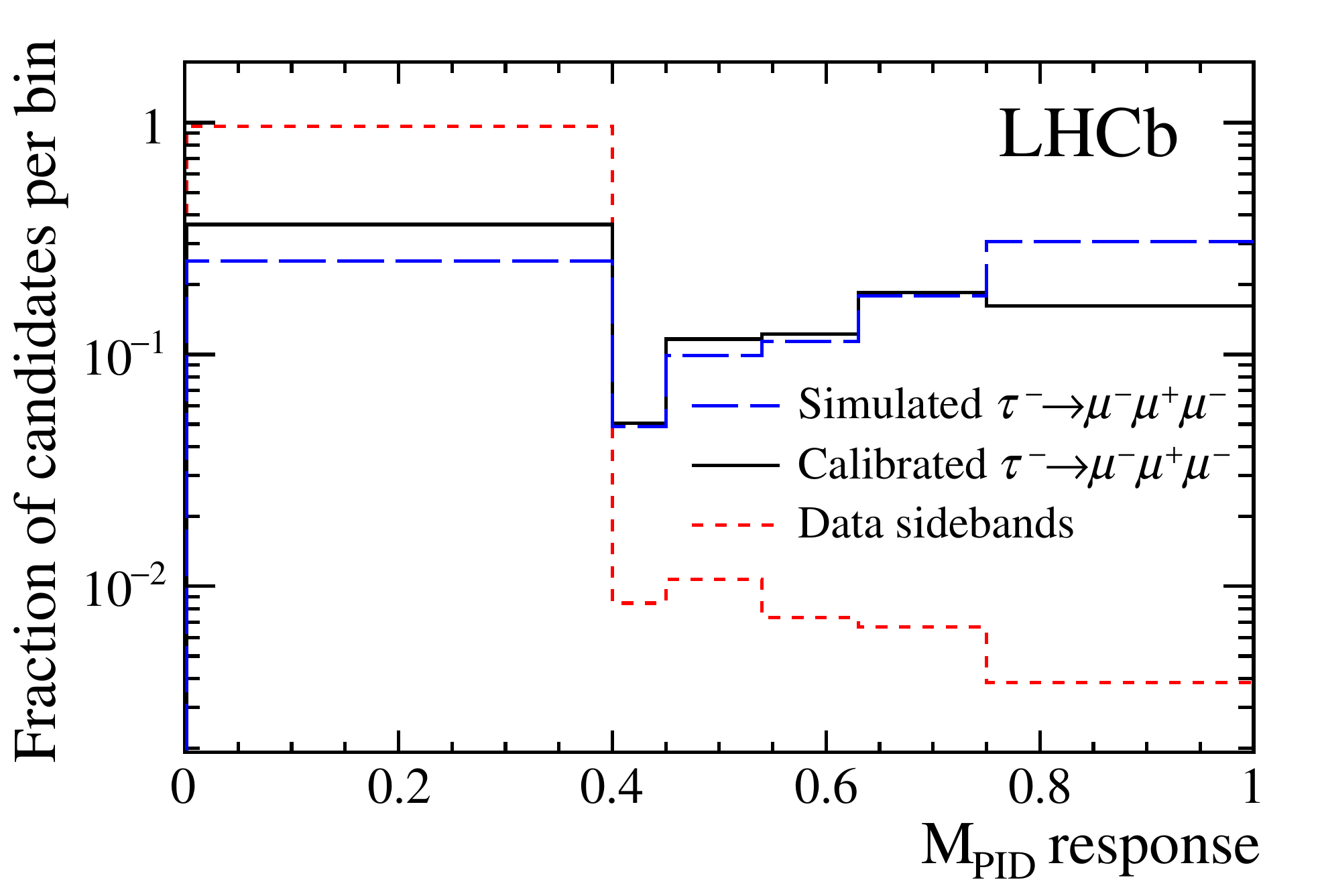}
        \put (40,115) {\small{(b)}}
        \end{overpic}
\end{minipage}
\begin{minipage}[b]{0.5\linewidth}
        \centering
        \begin{overpic}[width=0.95\textwidth]{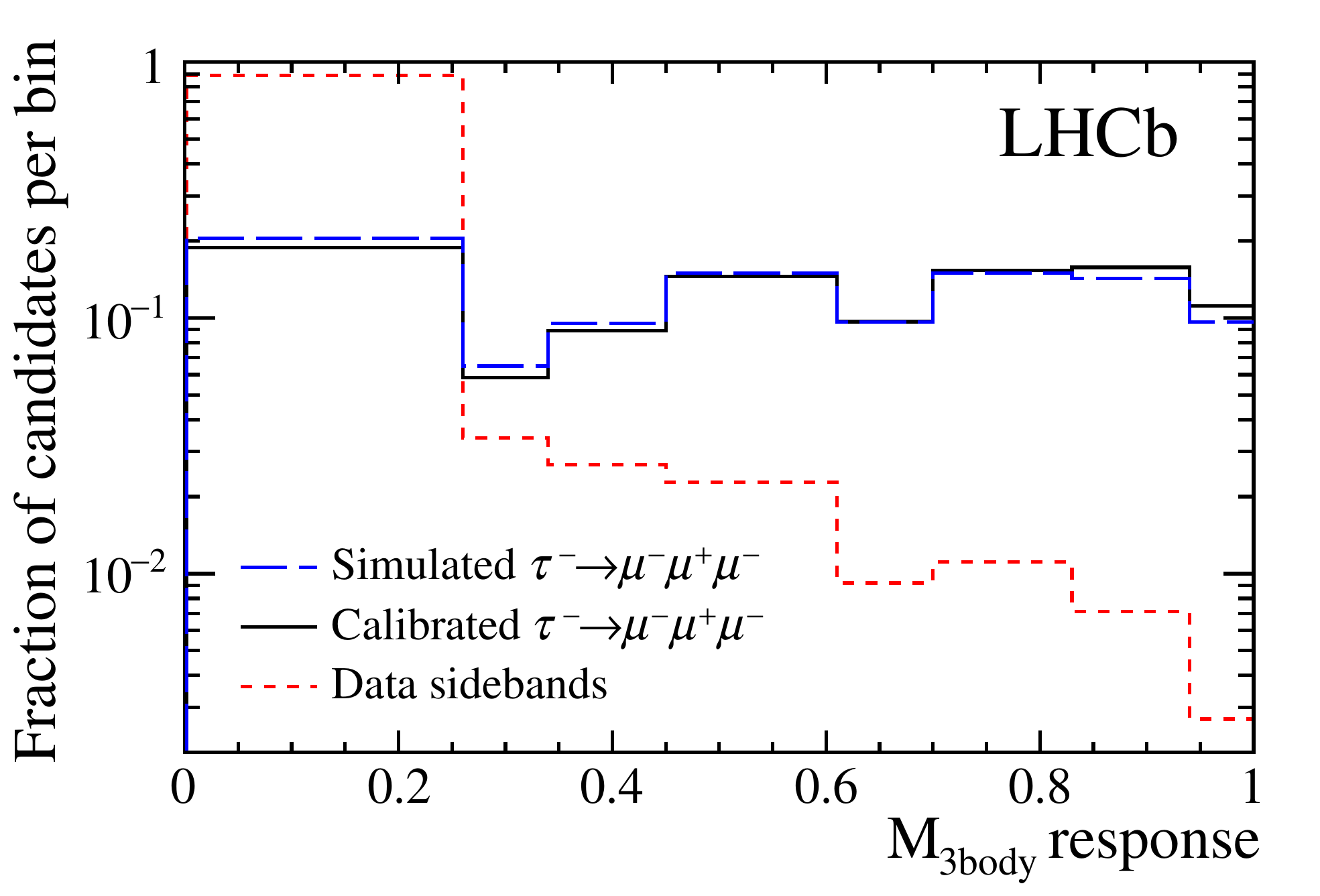}
        \put (40,117) {\small{(c)}}
        \end{overpic}
\end{minipage}
\begin{minipage}[b]{0.5\linewidth}
        \centering
        \begin{overpic}[width=0.95\textwidth]{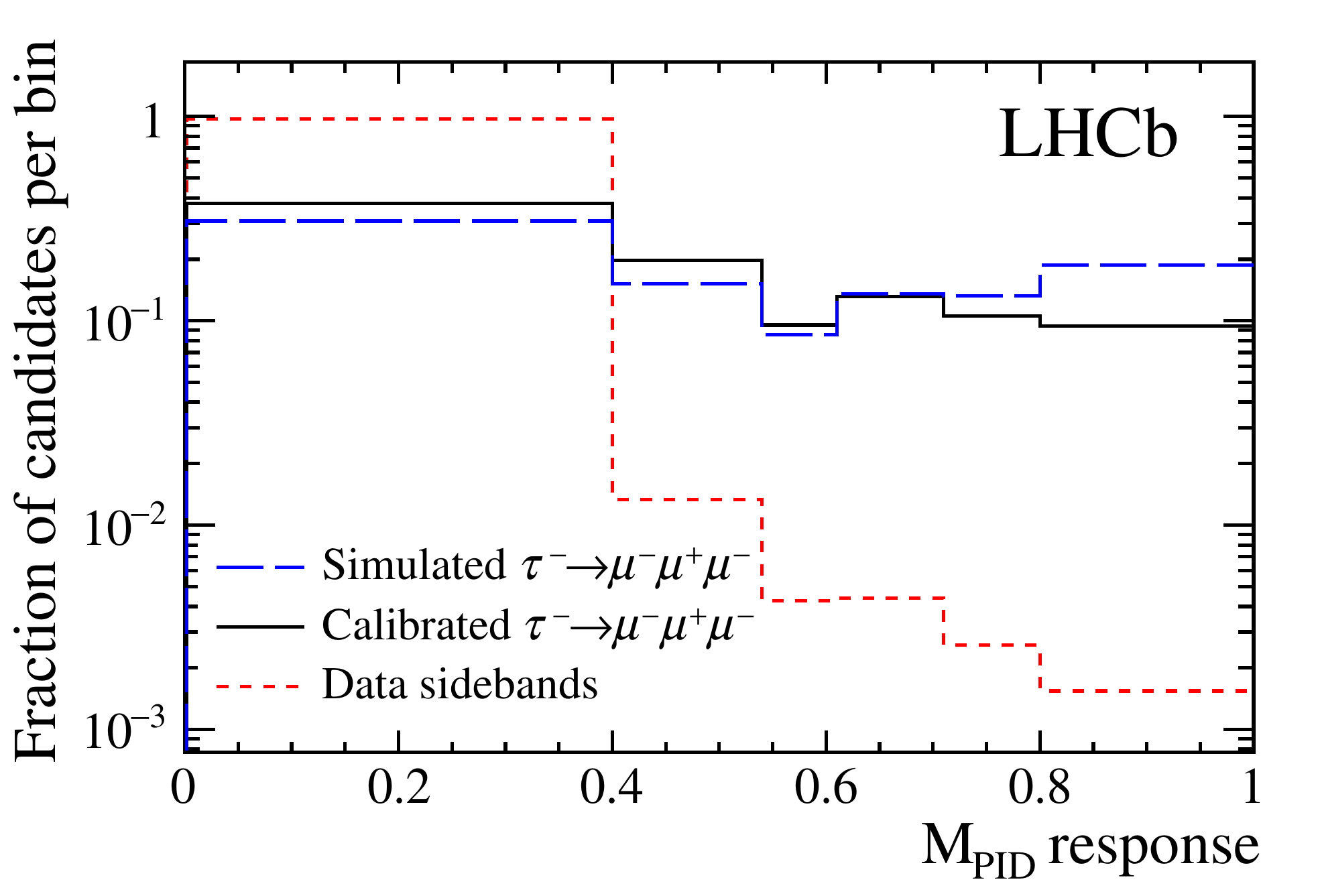}
        \put (40,117) {\small{(d)}}
        \end{overpic}
\end{minipage}

\caption{\small {Distribution of (a) \gl and (b) \pid response for 7\tev data and (c) \gl and (d) \pid response for 8\tev data. 
The binnings correspond to those used in the 
extraction of the final results. 
The short-dashed (red) lines show the response of the data sidebands, whilst the long-dashed (blue) and solid (black) lines show the response of 
simulated signal events before and after calibration.
In all cases the first bin is excluded from the analysis.}}
\label{fig:3muPDF}
\end{figure}

For the \gl and \pid responses, 
a binning is chosen via the \CLs method~\cite{Read_02, *Junk_99} by maximising the difference between 
the median \CLs values under the background-only hypothesis 
and the signal-plus-background hypothesis, whilst minimising the number of bins. 
The binning optimisation is performed separately 
for the 7\tev and 8\tev data sets, because there are small differences in event topology with 
changes of centre-of-mass energy. The optimisation does not depend on the signal branching fraction.
The bins at lowest values of \gl and \pid response do not
contribute to the sensitivity and are excluded from the analysis.
The distributions of the responses of the two classifiers, along with their binning schemes, are shown in Fig.~\ref{fig:3muPDF}. 

\begin{figure}[t]
        \centering
        \includegraphics[width=0.95\textwidth]{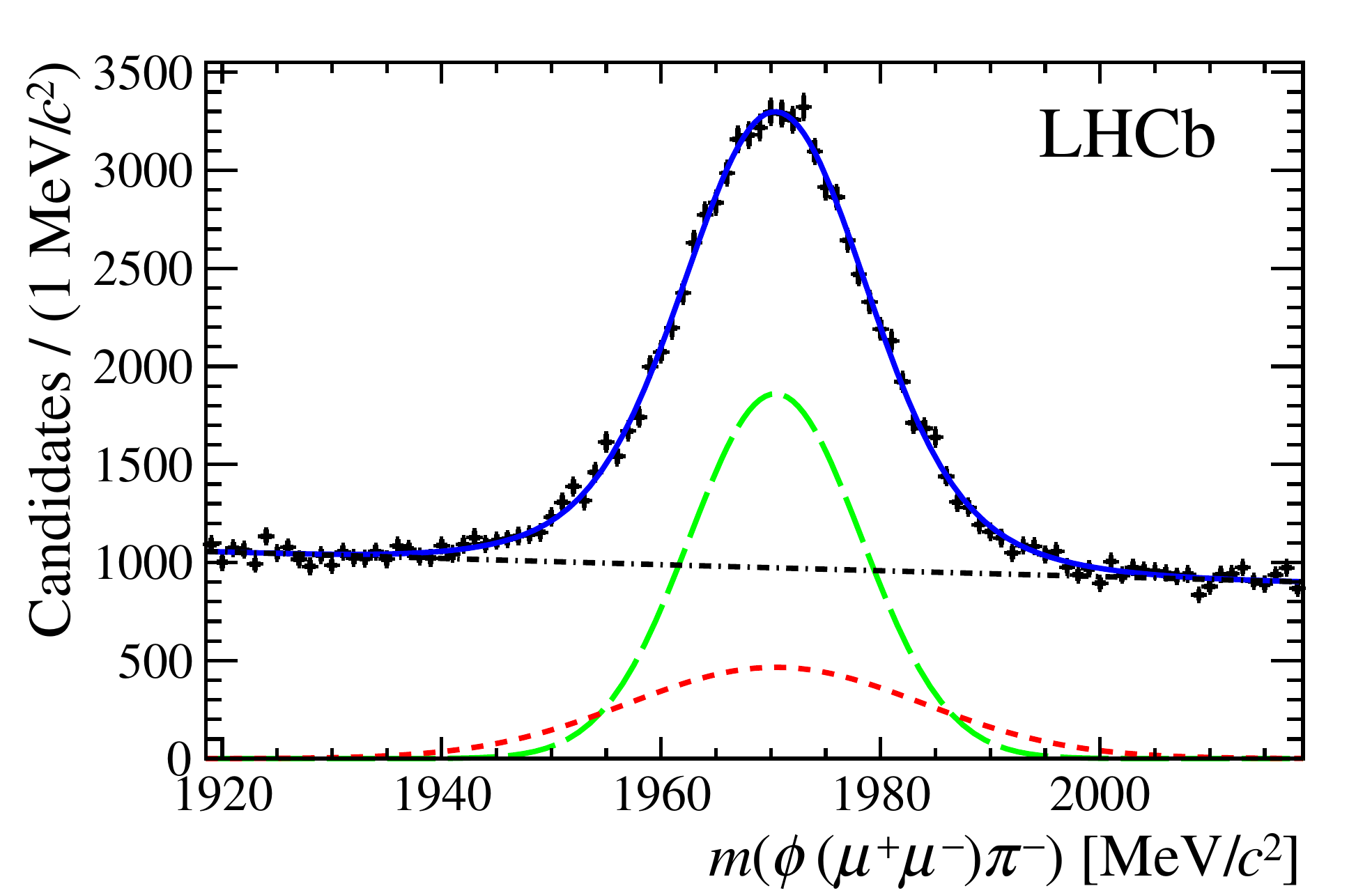}
\caption{\small {Invariant mass distribution of $\phi(\mu^+\mu^-)\pi^-$ candidates in 8\tev data. 
The solid (blue) line shows the overall fit, the long-dashed (green) and short-dashed (red) lines show the two Gaussian 
components of the $D^-_s$ signal and the dot-dashed (black) line shows the combinatorial background contribution.}}  
\label{fig:num_Ds}
\end{figure}

The expected shapes of the invariant mass spectra for the \tmmm signal in the 7\tev
and 8\tev data sets are taken from fits to the \DsPhiPi control 
channel in data. Figure~\ref{fig:num_Ds} shows the fit to the 8\tev data. 
No particle identification requirements are applied to the pion.
The signal distribution is modelled with 
the sum of two Gaussian functions with a common mean, where the narrower Gaussian   
contributes 70\% of the total signal yield, while the combinatorial background is modelled 
with an exponential function. The expected width of the $\tau^-$ signal in data is taken from 
simulation, scaled by the ratio of the widths of the $D_s^-$ peaks in data and simulation. 

\section{Backgrounds}
\label{sec:backgrounds}

The background processes for the \tmmm decay consist mainly of heavy meson decays yielding
three muons in the final state, or one or two muons in combination with two or one misidentified
particles. There are also a large number of events with one or two muons from heavy meson decays combined with 
two or one muons from elsewhere in the event.
Decays containing undetected final-state particles, such as \KL mesons, neutrinos or photons, 
can give large backgrounds, which vary smoothly in the signal region.
The most important background channel of this type is found to be 
\DsEtaMuNu, about $90\%$ of which is removed by the requirement on the dimuon mass.
The small remaining contribution from this process has a mass distribution similar to 
that of the other backgrounds in the mass range considered in the fit.
The dominant contributions to the background from misidentified particles
are from $D_{(s)}^-\rightarrow K^+\pi^-\pi^-$ and $D_{(s)}^-\rightarrow \pi^+\pi^-\pi^-$ decays.
However, these events populate mainly the region of low \pid response and are reduced 
to a negligible level by the exclusion of the first bin.

\begin{figure}[t]
\begin{minipage}[b]{0.5\linewidth}
	\centering
	\begin{overpic}[width=\textwidth]{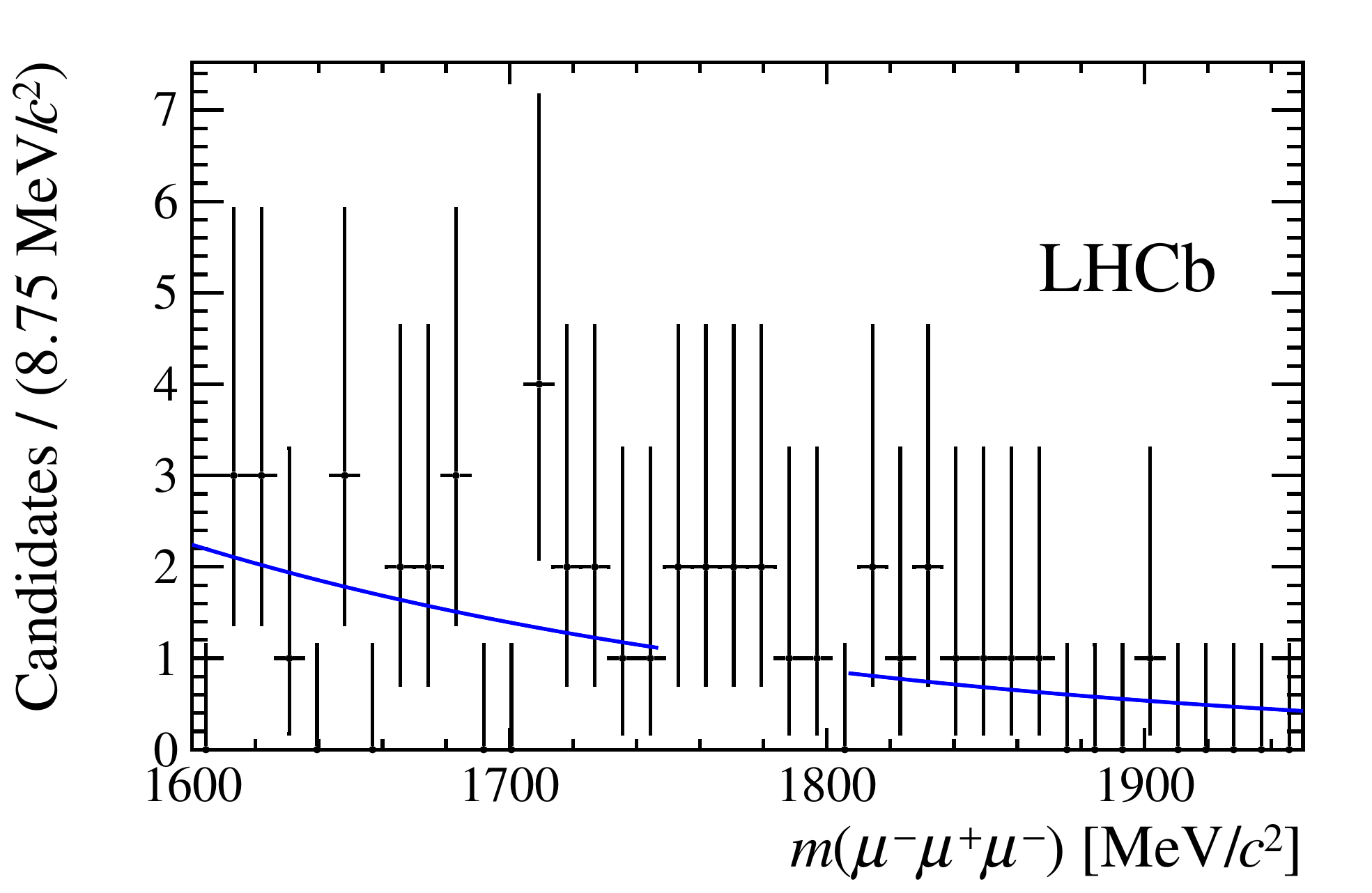}
	\put (50,120) {\small{(a)}}
	\put (117,125) {\footnotesize{\gl$\in [0.80, 1.0]$}}
	\put (117,115) {\footnotesize{\pid$\in [0.75, 1.0]$}}
	\end{overpic}
\end{minipage}
\begin{minipage}[b]{0.5\linewidth}
	\centering
	\begin{overpic}[width=\textwidth]{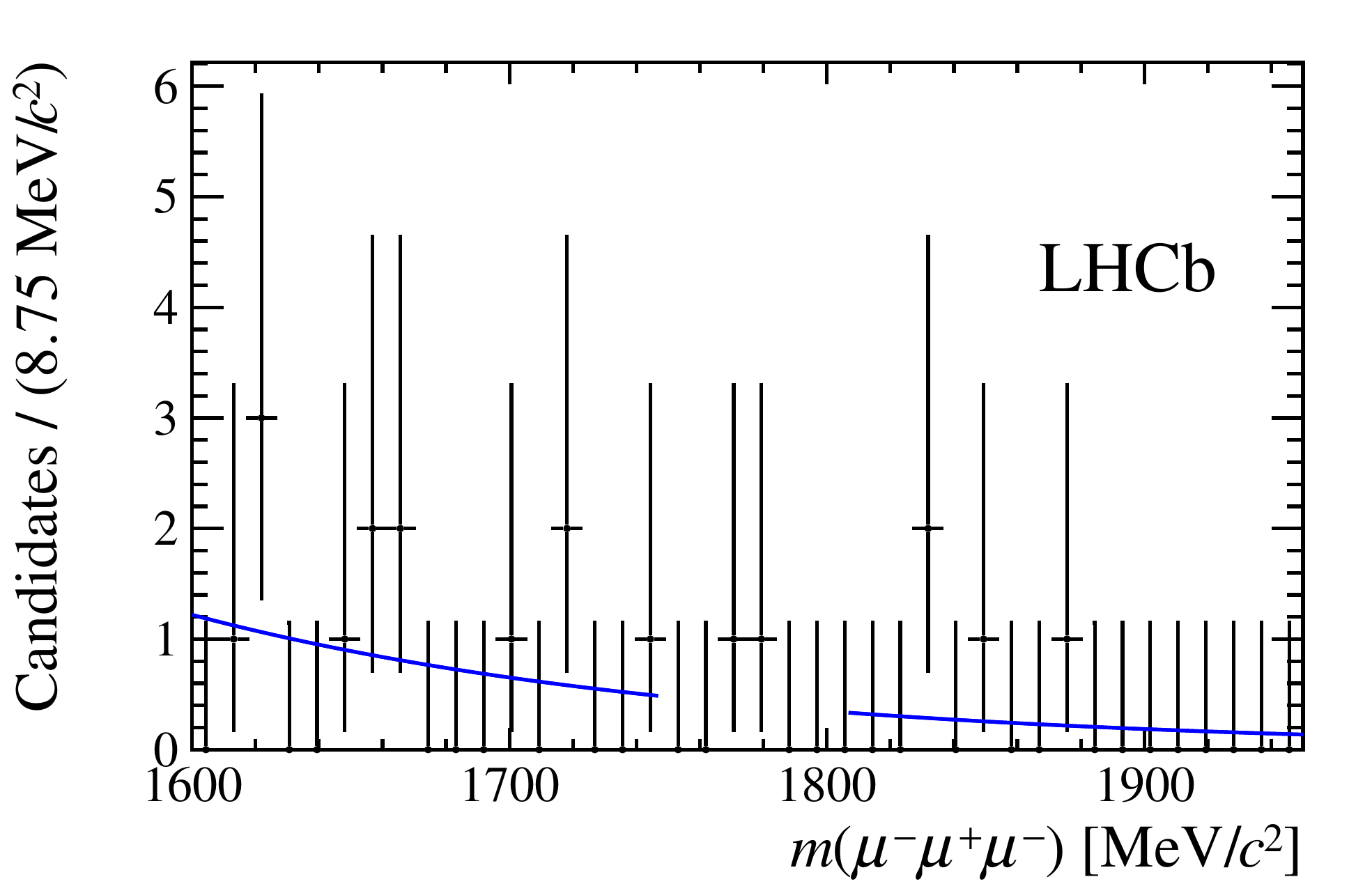}
	\put (50,120) {\small{(b)}}
	\put (117,125) {\footnotesize{\gl$\in [0.94, 1.0]$}}
	\put (117,115) {\footnotesize{\pid$\in [0.80, 1.0]$}}
	\end{overpic}
\end{minipage}
\caption
{\small{Invariant mass distributions and fits to the mass sidebands in (a) 7\tev and (b) 8\tev data for $\mu^+\mu^-\mu^-$ 
candidates in the bins of \gl and \pid response that contain the highest signal probabilities.}}
  \label{fig:bkg_fits}
\end{figure}
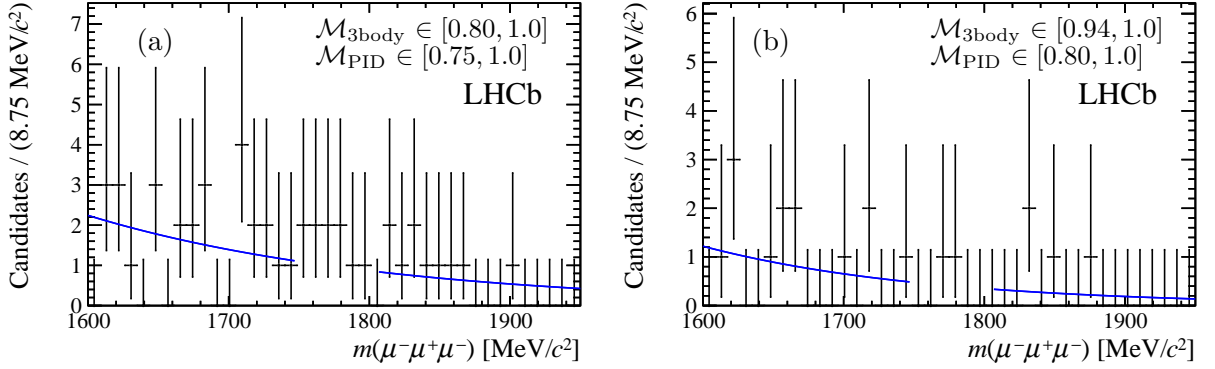

The expected numbers of background events within the signal region, for each bin in 
\gl and \pid,
are evaluated by fitting an exponential function to the candidate mass spectra outside of the signal windows 
using an extended, unbinned maximum likelihood fit. 
The parameters of the exponential function are allowed to vary independently in each bin.
The small differences obtained if the exponential curves are 
replaced by straight lines are included as systematic uncertainties.
The $\mu^-\mu^+\mu^-$ mass spectra are fitted over the
mass range 1600--1950\mevcc, excluding windows of width $\pm 30$\mevcc around the expected signal mass. 
The resulting fits to the data sidebands for the highest sensitivity bins are shown in Fig.~\ref{fig:bkg_fits} for 7 and 8\tev 
data separately.  

\section{Normalisation}
\label{sec:norm}

The observed number of \tmmm candidates is converted into a branching fraction by normalising 
to the \DsPhiPi calibration channel 
according to
\begin{equation}
{\BRof\tmmm}=\frac{\BRof\DsPhiPi}{\BRof\DsTauNu}
\times
f^{D_{s}}_{\tau}
\times
\frac { \rm {\epsilon\mathstrut_{cal}^{R}}}
{\rm \epsilon\mathstrut_{sig}^{R}}
\times
\frac{\rm {\epsilon\mathstrut_{cal}^{T}}}
{\rm \epsilon\mathstrut_{sig}^{T}}
\times\frac{N_{\rm sig}}{N_{\rm cal}} 
\equiv \alpha N_{\rm sig}\, ,
\label{eq:norm}
\end{equation}
where $\alpha$ is the overall normalisation factor, $N_{\rm sig}$ is the
number of observed signal events and all other terms are described below. 
Table~\ref{tab:norm_summary} gives a summary of all contributions to the factor
$\alpha$; the uncertainties are taken to be uncorrelated.
The branching fraction of the normalisation channel 
is determined from known branching fractions as
\begin{equation*}
\BRof{ \DsPhiPi } = \frac{ \BRof{ \DsPhiKKPi} } 
                           {  \BRof{ \PhiKK } } 
                    \BRof{ \Phimm }=\left(1.32\pm0.10\right)\times 10^{-5}\, ,
\end{equation*} 
\noindent where \BRof{ \PhiKK } and 
\BRof{ \Phimm } are taken from Ref.~\cite{PDG2014} and $\BRof{ \DsPhiKKPi}$ is taken from Ref.~\cite{Babar2011a}.
The branching fraction \BRof\DsTauNu is 
taken from Refs.~\cite{PDG2014,Belle_Ds_branchingfractions}. 

The quantity $f^{D_{s}}_{\tau}$ is the
fraction of $\tau^-$ leptons that originate from $D_s^-$ decays. 
The value of $f^{D_{s}}_{\tau}$ at 7\tev is calculated using the
$b\bar{b}$ and $c\bar{c}$ cross-sections as measured by
\lhcb~\cite{sigmabbLHCb,sigmaccLHCb} at 7\tev and the inclusive 
$b\ra D_{s}$, $c\ra D_{s}$, 
$b\ra\tau$ and $c\ra\tau$ 
branching fractions~\cite{PDG2014}.
For the value of $f^{D_{s}}_{\tau}$ at 8\tev the $b\bar{b}$ cross-section is updated to the 
8\tev \lhcb measurement~\cite{bCrossSec8TeV} and the $c\bar{c}$ cross-section measured at 7\tev is 
scaled by a factor of \nicefrac{8}{7}, consistent with \pythia simulations. 
The uncertainty on this scaling factor, which is negligible, is found by taking the difference between the 
value obtained from the nominal parton distribution functions and that from the average of their corresponding error sets~\cite{CTEQ}.

The reconstruction and selection efficiencies, $\rm \epsilon^{R}$, are
products of the detector acceptances for the decay of interest, the muon identification efficiencies 
and the selection efficiencies. 
The combined muon identification and selection efficiencies are determined from the
yield of simulated events after the full selections are applied. 
The ratio of efficiencies is corrected to account for the differences between data and simulation 
in track reconstruction, muon identification, the $\phi(1020)$ mass window 
requirement in the normalisation channel 
and the $\tau^-$ mass range.
The removal of candidates in the least sensitive bins in the \gl and \pid classifier responses
is also taken into account. 

The trigger efficiencies, $\rm \epsilon^{T}$, are
evaluated from simulation and their systematic uncertainties are determined from
the differences between the trigger efficiencies of \BuJpsimmK decays
measured in data and in simulation, using muons with momentum values typical of \tmmm signal decays. 
The trigger efficiency for the 8\tev data set is 
corrected to account for differences in trigger conditions across the data taking period, 
resulting in a relatively large systematic error.

The yields of \DsPhiPi candidates in data,
$N_{\rm cal}$, are determined from the fits to reconstructed $\phi\left(\mu^+\mu^-\right)\pi^-$ mass distributions 
shown in Fig.~\ref{fig:num_Ds}. 
The variations in the yields when the relative 
contributions of the two Gaussian components are allowed to vary in the fits are considered as systematic uncertainties. 

\renewcommand*\arraystretch{1.5}
\begin{table}[t]
\centering
\caption[]{\small {Terms entering into the normalisation factors, $\alpha$, 
and their combined statistical and systematic uncertainties.}}
\vspace{1mm}
\label{tab:norm_summary}
$
\begin{array}{cr@{\hspace{1mm}\pm\hspace{1mm}}lr@{\hspace{1mm}\pm\hspace{1mm}}l}
\toprule
& \multicolumn{2}{c}{7\tev} & \multicolumn{2}{c}{8\tev} \\   
\midrule
\BRof\DsPhiPi & \multicolumn{2}{r@{\hspace{1mm}\pm\hspace{1mm}}}{\left(1.32\right.} & \multicolumn{2}{@{}l}{\left.0.10\right) \times 10^{-5}}\\
\BRof\DsTauNu & \multicolumn{2}{r@{\hspace{1mm}\pm\hspace{1mm}}}{\left(5.61\right.} & \multicolumn{2}{@{}l}{\left.0.24\right) \times 10^{-2}}\\
f^{D_{s}}_{\tau} & 0.78 & 0.04 & 0.80 & 0.03\\
\rm{\epsilon\mathstrut_{cal}}^{R}/
\rm{\epsilon\mathstrut_{sig}}^{R}
& 0.898 & 0.060 & 0.912 & 0.054\\  
\rm{\epsilon\mathstrut_{cal}}^{T}/
\rm{\epsilon\mathstrut_{sig}}^{T}  
& 0.659 & 0.006 & 0.525 & 0.040\\  
N_{\rm cal} & 28\,200 & 440 & 52\,130 & 700\\
\alpha & \left( 7.20 \right. & \left. 0.98 \right) \times 10^{-9} & \left( 3.37 \right. & \left. 0.50 \right) \times 10^{-9}\\
\bottomrule
\end{array}
$
\end{table}
\renewcommand*\arraystretch{1}

\section{Results}
\label{sec:results}

\begin{figure}[h]
\begin{center} 
	\includegraphics[width=\textwidth]{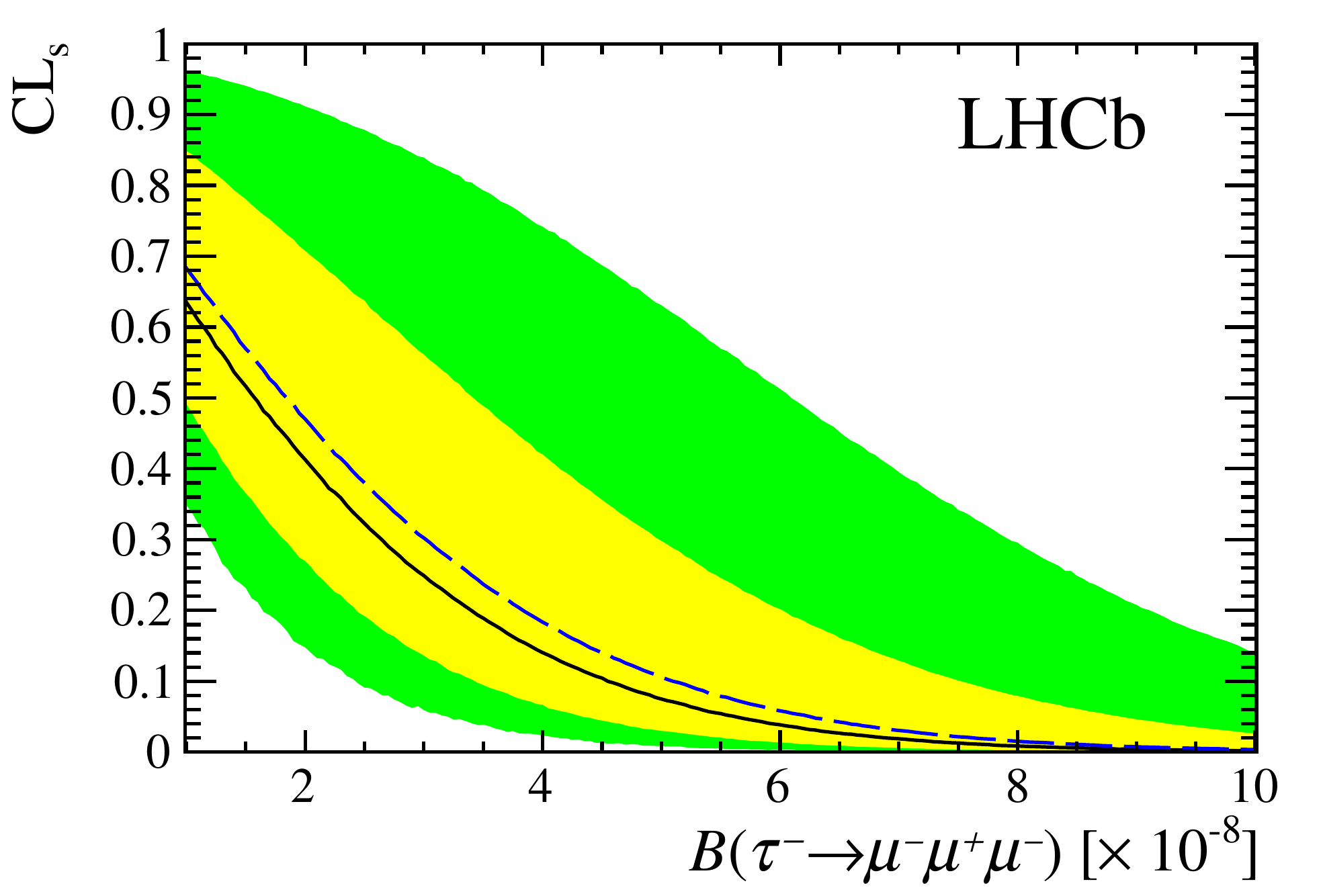}
\caption
{\small {Distribution of \CLs values as a function of the assumed
  branching fraction for \tmmm, under the hypothesis to observe background events only. 
The dashed line indicates the expected limit and the
  solid line the observed one.
  The light (yellow) and dark (green) bands cover the regions of 68\% and 95\%
  confidence for the expected limit.}
}
\label{fig:CLs}
\end{center} 
\end{figure}

Tables~\ref{tab:bkg_tmmm2011} and~\ref{tab:bkg_tmmm2012} give the expected and observed numbers of candidates in the signal region, for each bin 
of the classifier responses. No significant excess of events over the expected background is observed.
Using the \CLs method~\cite{Read_02, *Junk_99} and Eq.~\ref{eq:norm},
the observed \CLs value and the expected \CLs distribution are calculated as functions of the assumed branching fraction, 
as shown in Fig.~\ref{fig:CLs}.
The systematic uncertainties on the signal and background estimates, which have a very small effect on the final limits, 
are included following Ref.~\cite{Read_02, *Junk_99}.
The expected limit at $90\%~(95\%)$ CL for the branching
fraction is $\BRof\tmmm < 5.0~(6.1) \times 10^{-8}$, while the observed limit at $90\%~(95\%)$ CL is
\begin{center}
$\BRof\tmmm < 4.6~(5.6) \times 10^{-8}.$
\end{center}
Whilst the above limits are given for the phase-space model of $\tau^-$ decays, the kinematic properties of the decay 
would depend on the physical processes that introduce LFV. 
Reference~\cite{LFVmodels} gives a model-independent analysis of the decay distributions in an effective field-theory approach 
including
BSM operators with different chirality structures. 
Depending on the choice of operator, the observed limit varies within the range $(4.1 - 6.8) \times 10^{-8}$ 
at $90\%$ CL. 
The weakest limit results from an operator that favours low $\mu^+\mu^-$ mass, since the requirement
to remove the 
\DsEtaMuNu\, 
background excludes a large fraction of the relevant phase-space.  

In summary, the LHCb search for the LFV decay \tmmm is updated using all
data collected during the first run of the LHC, corresponding to an integrated luminosity of 3.0\invfb. 
No evidence for any signal is found. 
The measured limits supersede those of Ref.~\cite{paper1fb} and, in combination with results from the \textit{B} factories, improve the constraints placed on the parameters of a broad class of BSM models~\cite{HFAG_2014}.

\begin{table}[p]
\begin{center}
\caption{\small {Expected background candidate yields in the 7\tev data set, with their uncertainties, 
and observed candidate yields within the $\tau^-$ signal window in the different bins of classifier response. 
The classifier responses range from
$0$ (most background-like) to $+1$ (most signal-like).
The first
bin in each classifier response is excluded from the analysis.}}
\small
\label{tab:bkg_tmmm2011}
\begin{tabular}{cr@{\hspace{1mm}--\hspace{1mm}}lr@{\hspace{1mm}$\pm$\hspace{1mm}}lc}
\toprule
\pid response & \multicolumn{2}{c}{\gl response} & \multicolumn{2}{c}{Expected} & Observed\\
\midrule
 & 0.28 & 0.32 & 3.17 & 0.66 & 4\\
 & 0.32 & 0.46 & 9.2 & 1.1 & 6\\
 0.40 -- 0.45 & 0.46 & 0.54 & 2.89 & 0.63 & 6\\
 & 0.54 & 0.65 & 3.17 & 0.66 & 4\\
 & 0.65 & 0.80 & 3.64 & 0.72 & 2\\
 & 0.80 & 1.00 & 3.79 & 0.80 & 3\\
\midrule
 & 0.28 & 0.32 & 4.22 & 0.78 & 6\\
 & 0.32 & 0.46 & 8.3 & 1.1 & 10\\
 0.45 -- 0.54 & 0.46 & 0.54 & 2.3 & 0.57 & 4\\
 & 0.54 & 0.65 & 2.83 & 0.63 & 8\\
 & 0.65 & 0.80 & 2.72 & 0.69 & 5\\
 & 0.80 & 1.00 & 4.83 & 0.90 & 7\\
\midrule
 & 0.28 & 0.32 & 2.33 & 0.58 & 6\\
 & 0.32 & 0.46 & 8.3 & 1.1 & 8\\
 0.54 -- 0.63 & 0.46 & 0.54 & 2.07 & 0.53 & 1\\
 & 0.54 & 0.65 & 3.29 & 0.68 & 1\\
 & 0.65 & 0.80 & 2.96 & 0.65 & 4\\
 & 0.80 & 1.00 & 3.11 & 0.69 & 3\\
\midrule
 & 0.28 & 0.32 & 2.69 & 0.62 & 1\\
 & 0.32 & 0.46 & 7.5 & 1.0 & 5\\
 0.63 -- 0.75 & 0.46 & 0.54 & 2.06 & 0.53 & 3\\
 & 0.54 & 0.65 & 2.00 & 0.55 & 5\\
 & 0.65 & 0.80 & 3.16 & 0.66 & 2\\
 & 0.80 & 1.00 & 4.67 & 0.84 & 2\\
\midrule
 & 0.28 & 0.32 & 2.19 & 0.55 & 2\\
 & 0.32 & 0.46 & 3.38 & 0.76 & 5\\
 0.75 -- 1.00 & 0.46 & 0.54 & 1.52 & 0.46 & 3\\
 & 0.54 & 0.65 & 1.28 & 0.47 & 1\\
 & 0.65 & 0.80 & 2.78 & 0.65 & 1\\
 & 0.80 & 1.00 & 4.42 & 0.83 & 7\\
\bottomrule
\end{tabular} 
\end{center}
\end{table}

\begin{table}[p]
\begin{center}
\caption{\small {Expected background candidate yields in the 8\tev data set, with their uncertainties, 
and observed candidate yields within the $\tau^-$ signal window in the different bins of classifier response. 
The classifier responses range from
$0$ (most background-like) to $+1$ (most signal-like).
The first
bin in each classifier response is excluded from the analysis.}}
\small
\label{tab:bkg_tmmm2012}
\begin{tabular}{cr@{\hspace{1mm}--\hspace{1mm}}lr@{\hspace{1mm}$\pm$\hspace{1mm}}lc}
\toprule
\pid response & \multicolumn{2}{c}{\gl response} & \multicolumn{2}{c}{Expected} & Observed\\
\midrule
 & 0.26 & 0.34 & 39.6 & 2.3 & 39\\
 & 0.34 & 0.45 & 32.2 & 2.1 & 34\\
 & 0.45 & 0.61 & 28.7 & 2.0 & 28\\
 0.40 -- 0.54 & 0.61 & 0.70 & 9.7 & 1.2 & 5\\
 & 0.70 & 0.83 & 11.4 & 1.3 & 7\\
 & 0.83 & 0.94 & 7.3 & 1.1 & 6\\
 & 0.94 & 1.00 & 6.0 & 1.0 & 0\\
\midrule
 & 0.26 & 0.34 & 13.6 & 1.4 & 8\\
 & 0.34 & 0.45 & 12.1 & 1.3 & 12\\
 & 0.45 & 0.61 & 8.3 & 1.0 & 13\\
 0.54 -- 0.61 & 0.61 & 0.70 & 2.60 & 0.62 & 1\\
 & 0.70 & 0.83 & 1.83 & 0.60 & 5\\
 & 0.83 & 0.94 & 2.93 & 0.72 & 6\\
 & 0.94 & 1.00 & 2.69 & 0.63 & 3\\
\midrule
 & 0.26 & 0.34 & 13.5 & 1.4 & 7\\
 & 0.34 & 0.45 & 10.9 & 1.2 & 11\\
 & 0.45 & 0.61 & 9.7 & 1.2 & 12\\
 0.61 -- 0.71 & 0.61 & 0.70 & 3.35 & 0.69 & 2\\
 & 0.70 & 0.83 & 4.60 & 0.89 & 5\\
 & 0.83 & 0.94 & 4.09 & 0.81 & 4\\
 & 0.94 & 1.00 & 2.78 & 0.68 & 1\\
\midrule
 & 0.26 & 0.34 & 7.8 & 1.1 & 6\\
 & 0.34 & 0.45 & 7.00 & 0.99 & 8\\
 & 0.45 & 0.61 & 6.17 & 0.95 & 6\\
 0.71 -- 0.80 & 0.61 & 0.70 & 1.57 & 0.56 & 2\\
 & 0.70 & 0.83 & 2.99 & 0.72 & 0\\
 & 0.83 & 0.94 & 3.93 & 0.81 & 0\\
 & 0.94 & 1.00 & 3.22 & 0.68 & 1\\
\midrule
 & 0.26 & 0.34 & 5.12 & 0.86 & 3\\
 & 0.34 & 0.45 & 4.44 & 0.79 & 6\\
 & 0.45 & 0.61 & 3.80 & 0.78 & 5\\
 0.80 -- 1.00 & 0.61 & 0.70 & 2.65 & 0.68 & 2\\
 & 0.70 & 0.83 & 3.05 & 0.67 & 2\\
 & 0.83 & 0.94 & 1.74 & 0.54 & 2\\
 & 0.94 & 1.00 & 3.36 & 0.70 & 3\\
\bottomrule
\end{tabular} 
\end{center}
\end{table}

\clearpage
\section*{Acknowledgements}

\noindent We express our gratitude to our colleagues in the CERN
accelerator departments for the excellent performance of the LHC. We
thank the technical and administrative staff at the LHCb
institutes. We acknowledge support from CERN and from the national
agencies: CAPES, CNPq, FAPERJ and FINEP (Brazil); NSFC (China);
CNRS/IN2P3 (France); BMBF, DFG, HGF and MPG (Germany); SFI (Ireland); INFN (Italy);
FOM and NWO (The Netherlands); MNiSW and NCN (Poland); MEN/IFA (Romania);
MinES and FANO (Russia); MinECo (Spain); SNSF and SER (Switzerland);
NASU (Ukraine); STFC (United Kingdom); NSF (USA).
The Tier1 computing centres are supported by IN2P3 (France), KIT and BMBF
(Germany), INFN (Italy), NWO and SURF (The Netherlands), PIC (Spain), GridPP
(United Kingdom).
We are indebted to the communities behind the multiple open
source software packages on which we depend. We are also thankful for the
computing resources and the access to software R\&D tools provided by Yandex LLC (Russia).
Individual groups or members have received support from
EPLANET, Marie Sk\l{}odowska-Curie Actions and ERC (European Union),
Conseil g\'{e}n\'{e}ral de Haute-Savoie, Labex ENIGMASS and OCEVU,
R\'{e}gion Auvergne (France), RFBR (Russia), XuntaGal and GENCAT (Spain), Royal Society and Royal
Commission for the Exhibition of 1851 (United Kingdom).

\addcontentsline{toc}{section}{References}
\bibliographystyle{LHCb}
\bibliography{main}

\newpage 

\centerline{\large\bf LHCb collaboration}
\begin{flushleft}
\small
R.~Aaij$^{41}$, 
B.~Adeva$^{37}$, 
M.~Adinolfi$^{46}$, 
A.~Affolder$^{52}$, 
Z.~Ajaltouni$^{5}$, 
S.~Akar$^{6}$, 
J.~Albrecht$^{9}$, 
F.~Alessio$^{38}$, 
M.~Alexander$^{51}$, 
S.~Ali$^{41}$
% ...

{\footnotesize \it
% ...
$ ^{5}$Clermont Universit\'{e}, Universit\'{e} Blaise Pascal, CNRS/IN2P3, LPC, Clermont-Ferrand, France\\
$ ^{6}$CPPM, Aix-Marseille Universit\'{e}, CNRS/IN2P3, Marseille, France\\
% ...
$ ^{9}$Fakult\"{a}t Physik, Technische Universit\"{a}t Dortmund, Dortmund, Germany\\
% ...
$ ^{37}$Universidad de Santiago de Compostela, Santiago de Compostela, Spain\\
$ ^{38}$European Organization for Nuclear Research (CERN), Geneva, Switzerland\\
% ...
$ ^{41}$Nikhef National Institute for Subatomic Physics, Amsterdam, The Netherlands\\
% ...
$ ^{46}$H.H. Wills Physics Laboratory, University of Bristol, Bristol, United Kingdom\\
% ...
$ ^{51}$School of Physics and Astronomy, University of Glasgow, Glasgow, United Kingdom\\
$ ^{52}$Oliver Lodge Laboratory, University of Liverpool, Liverpool, United Kingdom\\
% ...
}
\end{flushleft}
\end{document}